\documentclass[12pt]{emulateapj}
\def\deg{\hbox{$^\circ$}}
\newcommand{\etal}{{\it et al. }}
\newcommand{\asec}{$^{\prime\prime }$}
\newcommand{\amin}{$^\prime $}
\shortauthors{Wilking et al.}
\shorttitle{The Outflow from Haro~6-10}

\begin{document}

\title{A Proper Motion Study of the Haro~6-10 Outflow: Evidence for a Subarcsecond Binary}

\author{Bruce A. Wilking}
\affil{Department of Physics and Astronomy, University of Missouri-St. Louis}
\affil{1 University Boulevard, St.  Louis, MO 63121}
\email{bwilking@umsl.edu}
\author{Kevin B. Marvel}
\affil{American Astronomical Society}
\affil{2000 Florida Avenue, NW, Suite 400, Washington, DC 20009}
\email{marvel@aas.org}
\author{Mark J. Claussen}
\affil{National Radio Astronomy Observatory (NRAO) Array Operations Center}
\affil{P.O. Box 0}
\affil{1003 Lopezville Road, Socorro, NM 87801}
\email{mclausse@nrao.edu}
\author{Bradley M. Gerling}
\affil{Department of Physics and Astronomy, University of Missouri-St.  Louis}
\affil{1 University Boulevard, St.  Louis, MO 63121}
\email{bmg5333@truman.edu}
\author{Alwyn Wootten}
\affil{NRAO}
\affil{520 Edgemont Road, Charlottesville, VA 22903-2475}
\email{awootten@nrao.edu}
\and

\author{Erika Gibb}
\affil{Department of Physics and Astronomy, University of Missouri-St.  Louis}
\affil{1 University Boulevard, St.  Louis, MO 63121}
\email{gibbe@umsl.edu}

\begin{abstract}
We present single-dish and VLBI observations of an outburst of water maser emission from the young binary system
Haro~6-10.
Haro~6-10 lies in the Taurus molecular cloud and contains a visible T Tauri 
star with an infrared companion 1.3\asec\ north. 
Using the Very Long Baseline Array, we obtained five observations spanning 3 months and derived 
absolute positions for 20 distinct maser 
spots.  Three of the masers can be traced over 3 or more epochs, enabling us to 
extract absolute proper motions and tangential velocities.
We deduce that the masers represent one side of a bipolar outflow that lies nearly in the 
plane of the sky with an opening angle of
$\sim$45\deg.  They are located within 50 mas of  
the southern component of the binary, the visible T Tauri star Haro~6-10S.
The mean position angle on the sky of the maser
proper motions ($\sim$220\deg) suggests they are related to 
the previously observed giant Herbig-Haro (HH) flow which includes HH~410, HH~411, HH~412, and HH~184A-E.   
A previously observed HH jet and extended radio continuum emission (mean position angle of 
$\sim$190\deg) must also originate in the vicinity of Haro~6-10S and represent a second, distinct outflow
in this region.  We propose that a yet unobserved companion within 150 mas of Haro~6-10S 
is responsible for the giant HH/maser outflow while the visible star is associated with the HH jet. 
Despite the presence of H$_2$ emission in the spectrum of the northern component of the binary, Haro~6-10N, 
none of outflows/jets can be tied directly to this young stellar object.

\end{abstract}

\section{Introduction}

Haro~6-10 (R.A.(2000) 4$^h$ 29$^m$ 23.7$^s$, decl.(2000.0) +24$^{\circ}$ 33\amin\ 00\asec) 
is a young binary system containing a visible T Tauri star (Haro~6-10S also known as GV~Tau~S) 
and an infrared companion (Haro~6-10N also known as GV~Tau~N). 
They lie in the L~1534 molecular core in the Taurus cloud (140 pc, Elias 1978; Torres \etal 2009)
and are separated by 1.3\asec\ (180 AU) at a position angle of 355 degrees (Leinert \& Haas 1989). 
Both components exhibit 
late-type photospheres (K7-M2) with luminosities and masses on the order of the Sun's 
(Goodrich 1986, Doppmann \etal 2008).  Haro~6-10N, the more massive of the two objects, 
is apparently obscured by an edge-on disk and overtakes Haro~6-10S in brightness 
at wavelengths $\ge$4.8 $\mu$m (Leinart \& Haas 1989; Menard \etal 1993). 
Both components of the Haro~6-10 system are well-studied with high-resolution 
infrared spectroscopy.  Both show signatures of 
accretion and outflow through emission lines of Brackett~$\gamma$ and molecular hydrogen while 
Haro~6-10N is one of only a few YSOs which display absorption lines from CO and warm organic 
molecules such as HCN and C$_2$H$_2$ (Gibb \etal 2008; Doppmann \etal 2008).   
%Near-infrared monitoring by Leinert et al. (2001) over 12 years found that both
%stars showed significant variations: the decrease in flux of the southern component was
%attributed to increased extinction, while the increase in brightness of the northern source 
%could not be explained by lower extinction.
Recently, Roccatagliata et al. (2011) have presented high-resolution optical and near-infrared
images of Haro~6-10, as well as spectroscopic and interferometric observations in the mid-IR.
The results of these observations suggest that Haro~6-10N and Haro~6-10S are embedded
in a common envelope, each with a circumstellar disk-like structure. Moreover, the disks
are highly misaligned, with the disk around the northern component mostly edge-on and the disk
around Haro~6-10S almost face-on. 

Multiple outflows are known in the Haro~6-10 region.  The dominant outflow is a highly collimated
giant Herbig-Haro (HH) flow that extends for 1.6 pc and contains HH~411 and HH~412 (northeast) 
and HH~410 (southwest) at a position angle of $\sim$222\deg\ (Devine \etal 1999).  
This outflow was initially associated with Haro~6-10N because of that YSO's more highly obscured, and
presumably less evolved, nature.  Much closer to Haro~6-10 is the HH~184A-E group 
which lies about 3\amin\ (0.12 pc) to the southwest at position angles varying
from 225\deg\ to 249\deg.   The presence of this lightly obscured HH group led 
Devine \etal to suggest the southwestern flow is blue-shifted.  
Indeed, a low-velocity bipolar molecular outflow with blue-shifted gas to the southwest
and red-shifted gas to the northeast is observed (Stojimirovic, Narayanan, \& Snell 2007).
A second outflow with a north-south
orientation (PA = 175\deg) is suggested by the one-sided morphology of HH objects 184F-G
(Devine \etal 1999).
A similar orientation relative to Haro 6-10 is seen for a blue-shifted HH jet
observed in emission from [S~II], [O~I], [Fe~II], and Br$\gamma$ with position angles 
ranging from 175\deg\ to 195\deg\ (Movsessian \& Magakian 1999; Beck, Bary, \& McGregor 2010). 
High resolution radio continuum 
observations at $\lambda$=3.6 cm have shown a compact radio jet to the southwest 
of Haro~6-10S at a position angle of 191\deg\ that,
along with extended Br$\gamma$ emission, definitively ties the HH jet to the Haro~6-10S (Reipurth \etal 2004; Beck \etal 2010).  
Reipurth \etal propose that 
that precession of the outflow, possibly due to a close companion, has shifted the position angle from that
of the giant HH flow to that of the Herbig-Haro/radio jet.  
%A close companion to Haro~6-10S has been proposed based on the proposed precession and the presence 
%of a possible second radio jet (Reipurth \etal). 
%and apparent radial velocity variations 
%measured for the Haro~6-10S (Doppmann \etal 2008).

Water masers are highly variable tracers of energetic outflows from young stars.  
Shocks produced as the outflows sweep up ambient material enhance the water abundance and, 
more importantly, collisionally pump the water molecules to an excited state from 
which they produce stimulated emission.  While the first masers found were associated with 
young massive stars, water masers are also common around lower mass young stellar objects (YSOs), 
albeit with a lower duty cycle.  Hence it is necessary to monitor lower mass YSOs to 
detect episodes of maser emission.  
Single dish monitoring of intermediate-to-low mass YSOs has been used to trigger interferometric observations 
that track proper motions of water masers and trace their energetic winds 
(e.g., Claussen \etal 1998; Patel \etal 2000; Furuya \etal 2005;Imai \etal 2007;
Marvel \etal 2008).  Such multi-epoch observations, 
when separated by periods of weeks, can track individual masers from which one can 
derive their proper motions, space velocities, as well as the orientation of the wind.  
Moreover, the lack of extinction by dust at radio wavelengths allows one to observe the 
maser emission within tens of AUs of the wind origin.

We describe the first detection of water masers from the Haro~6-10 system in 2003 as 
first reported by Wootten \etal (2005).  Using both single dish and 
interferometric radio observations, we first detected and then monitored  
maser activity associated with Haro~6-10S. Absolute proper motions measured for three masers
allowed us to characterize the outflow 
in terms of the proper motions, space velocities, and position angles of the masers on the sky. 
In Sec. 2 we describe the observations and the data reduction.  
The locations and space velocities of the masers are described in Sec. 3.  
Their relationship to the young binary system and their link to the larger scale molecular 
outflow and giant HH flow are discussed in Sec. 4.  

\section{Observations and Data Reduction}

Observations were made using the Green Bank Telescope (GBT) and the Very Long Baseline Array (VLBA), 
both operated by the National Radio Astronomy Observatory (NRAO)
\footnote{The National Radio Astronomy Observatory is a facility of the National Science 
Foundation operated under cooperative agreement by Associated Universities, Inc.}.  
The single-dish and interferometric radio observations are described below.  
The telescopes and dates of observations are summarized in Table 1.  
   
%\documentclass{emulateapj}
%\begin{document}
\begin{deluxetable}{lcc}
\tablenum{1}
\tablewidth{0pt}
\tablecaption{Summary of Observations}
\tablehead{
\colhead{Telescope\tablenotemark{a}} & Epoch &  \colhead{Date}  \\
\colhead{} & &  (YYMMDD)
} 
\startdata
GBT   &   & 031022    \\
GBT   &   & 031025    \\
GBT   &   & 031109   \\
GBT   &   & 031114    \\
GBT   &   & 031229  \\
VLBA  & I & 040106    \\
VLBA  &II & 040123     \\
GBT   &   & 040129  \\
GBT   &   & 040202  \\
VLBA  &III& 040208   \\
VLBA  &IV & 040223  \\
GBT   &   & 040303 \\
VLBA  &V  & 040308 \\
VLBA  &VI & 040324 \\
GBT   &   & 040326 \\
GBT   &   & 040409 \\
\enddata
\tablenotetext{a}{Observations were conducted under proposals AGBT02A\_063 (GBT) and BC128 (VLBA).}
\end{deluxetable}

%\end{document}

\subsection{Green Bank Telescope}

Haro~6-10 was observed in 2003 October - 2004 April as part of a regular GBT monitoring program 
of water masers associated with YSOs.  Emission from the 6$_{16}\rightarrow$5$_{23}$ water maser transition at 22.235081 GHz 
was observed on dates given in Table 1. The 18-22 GHz K-band receiver was used, 
which has two beams at a fixed separation of 3\amin\ in azimuth and a beamwidth of about 36\asec. 
The spectrometer was configured to provide channel spacing corresponding to 0.04 km s$^{-1}$. 
Data were reduced using the NRAO's AIPS++ program.  Once maser activity was observed with the GBT, 
time was requested to begin observations with the VLBA.

\subsection{Very Long Baseline Array}

Stations of the VLBA and a single antenna from the Very Large Array were used on six dates beginning in 
2004 January and separated by about 16 days (Table 1). The pointing center was the 
radio continuum position for Haro~6-10S (VLA~1, Reipurth \etal 2004).  
The data from five epochs were reduced using the Astronomical Image Processing System (AIPS classic,
Greisen 2003) 
distributed and maintained by NRAO. The VLBA data were recorded with a 8 MHz bandwidth 
centered at a velocity of 11.0 km s$^{-1}$  relative to the local standard of rest (LSR).
The correlator mode provided a velocity resolution of 0.21 km s$^{-1}$ per channel.  
Observations of the phase reference source J0426+2327 with a 8MHz bandpass were conducted during each observing period.
Bandpass calibration was achieved through observations of 3C~84.
Self-calibration on the strongest maser emission was performed using standard methods to extend 
the dynamic range of the images. After all calibration solutions were determined and applied, 
the data were mapped using the AIPS task IMAGR.  Spectral cubes were formed with pixel 
cell sizes of 80 $\mu$arcseconds.  The beam size was typically 750 $\mu$arcsec $\times$ 290 $\mu$arcsec
with a position angle near 0\deg\ and the average rms noise was $\sim$5 mJy.
We were unable to use data from all of the stations for our 08 February observations and the resulting beam 
size is about 35\% larger. 
The absolute amplitude calibration is accurate to about 20\%.  

To retrieve the absolute positional information, 
the strongest maser was used as the phase reference to image the weak calibrator J0426+2327.  
Its position relative to the center of the phase-referenced map was equivalent to the position of the Haro~6-10S 
radio continuum position relative to the strongest maser component.  
The images of J0426+2327 from each epoch
were used to determine the offsets from the phase center in order to measure the absolute
position of the strongest maser.  
The peak flux density of the J0426+2327 varied from 43 - 92 mJy.

\section{Results}

Despite its inclusion in previous maser surveys, our detection is the first 
and only report of 
maser activity in the Haro~6-10 region (e.g., Felli, Palagi, \& Tofani 1992; Terebey, Vogel, \& Myers 1992;
Xiang \& Turner 1995;
Claussen \etal 1996; Furuya \etal 2003; Sunada \etal 2007).
Maser emission mapped with the VLBA has revealed a close asociation with Haro~6-10S. 

\subsection{The Radial Velocity of Haro~6-10S}

The ambient cloud core in which the  binary system resides has a v$_{LSR}$ of
6.4 km s$^{-1}$ with a weaker, contaminating cloud at 4 km s$^{-1}$ 
(Stojimirovic, Narayanan, \& Snell 2007).
Estimates for the radial velocity of Haro~6-10S span a range from -6.2 km s$^{-1}$ to 
9.4  km s$^{-1}$
and have led 
Doppmann \etal (2008) to suggest
it is a radial velocity variable.
Published values for v$_{LSR}$ include 3.1$\pm$3.8 km s$^{-1}$ in 1999 December, 
-6.2$\pm$1.5 km s$^{-1}$ in 2001 November, and 9.4 $\pm$1.7 km s$^{-1}$ in 2007 January (White and Hillenbrand
2004; Covey \etal 2006; Doppmann \etal 2008, respectively).  Doppmann \etal (2008) used these values to 
fit a model employing a binary companion with a mass of 0.13 M$_{\sun}$, a period of 
38 days and a separation from Haro~6-10S of 0.35 AU.

This proposed companion is discrepant with that proposed by Reipurth \etal (2004), who 
suggested the presence of an unseen companion with a separation of 38 AU based on a 
high resolution image obtained through maximum entropy deconvolution.  
The companion model supported by Reipurth (2000) links the formation of Herbig-Haro flows to the 
breakup and ejection of one component from a non-hierarchical triple 
system.  
%We will address how our data physically preclude these companion models below.

Fortunately, a high resolution infrared echelle spectrum of Haro~6-10S was taken in August of 2003
which was only 2.5 months prior to our GBT monitoring and 5 months prior to our first VLBA
observation.  This observation covered the CO lines from 4202-4267 cm$^{-1}$ and
4333-4397 cm$^{-1}$ and is described in detail by Gibb \etal (2007).   
Using a synthetic spectrum to match the observed CO lines, a radial velocity of 9.5$\pm$1.4 km s$^{-1}$
is derived.  
This value is consistent with previous measurements, except for that of Covey \etal, and casts
doubt on whether Haro~6-10S is a radial velocity variable.  
From here on, we will reference the radial velocities of the observed maser emission to 
9.5 km s$^{-1}$.

\subsection{GBT}

The first reported detection of water masers from Haro~6-10 occurred in 2003 October and 
multiple maser velocities were observed in its spectrum over the next 6 months.
These observations, which are shown in Fig.~1, bracketed the VLBA observations.  
The flux density of the strongest maser component at 11.3 km s$^{-1}$ varied from 15 Jy in 2003 October 
to a peak of 170 Jy in 2003 December.  By 2004 April, the feature had faded to less than 1 Jy.  
A second, broader feature at 8 km s$^{-1}$ also persisted throughout the monitoring period.  
Other velocities appeared briefly at 12 and 13 km s$^{-1}$ but had disappeared by 2003 December. 

\subsection{VLBA}

A summary of the maser positions and radial velocities observed from 2004 January through 2004 March
are presented in Table 2.  Two-dimensional gaussian fits were performed to the maser components 
in each map using the AIPS task
JMFIT.  For the first five epochs, from 4-7 distinct components were observed over a sufficient number of
velocity channels to permit a gaussian fit to the line profiles.  The maser emission was too weak to map in epoch VI.
All of the maser components are
within 50 milliarcseconds (mas) or 7 AU of the VLA position of Haro~6-10S.  Position offsets from the
$\lambda$ = 3.6 cm position of VLA~1 are given in Table 1 for the peak velocity channel and
are accurate to $\pm$0.5 pixels or $\pm$0.04 mas.  All but one of the maser components appear to be
red-shifted relative to Haro~6-10S with radial velocities near 11 km s$^{-1}$.  A distinctive and broad
8.3 km s$^{-1}$, 
also seen in the GBT spectra, appears to be slightly blue-shifted and lies 15 mas east from the main group.

Because of the proximity of the Taurus region, apparent maser positions must be corrected for variations due to parallax.  
Corrections derived for the parallactic modulation relative to the center of the parallactic ellipse 
are given in Table 3 for the five epochs of VLBA observations.   The largest corrections appear between Epochs I and II.
We note that Torres \etal (2009) have estimated a peculiar velocity  of 10.6 km s$^{-1}$ for Taurus sources based on VLBA observations of
four YSOs.  However, no adjustment was made for a possible drift of Haro~6-10S 
and the masers due to this motion which would result
in an offset of 32 mas to the southeast relative to the radio continuum source position measured in 2002.  
Based upon their positions, radial velocities, and velocity widths, we were able to track three of the maser
components over at least three epochs.  The absolute proper motions and the tangential velocities derived from
them are given in Table 4.
These proper motions have been corrected for the solar motion using the formulation of Abad \& Vieira (2005).
Using the solar motion relative to the LSR derived from Hipparcos data by Dehnen \& Binney (1998), 
(U$_0$, V$_0$, W$_0$) = (10.0, 5.25, 7.17) km s$^{-1}$, the corrections in RA and decl. are 
-10.0 mas yr$^{-1}$ and -5.30 mas yr$^{-1}$, respectively.  Corrections for Galactic rotation were 
on the order of 1 mas yr$^{-1}$ or less and not made.
The tangential velocities and their uncertainties were computed
from a linear least squares fit to the
motion over time.  Radial velocities in Table 4 are referenced to that of Haro~6-10S. 
%with the assumption 
%that the molecular material in which the maser shock travels was initially at rest with respect to the star.
The locations and position angles of the maser tangential velocities are shown in Fig. 2.

Since there is no evidence for a bipolar maser outflow
associated with this source as there is in others (e.g., IRAS~05413-0104, Claussen et al. 1998),
the outflow traced by the three masers apparently locates one side of a bipolar outflow with a 
velocity on the order of 10 km s$^{-1}$.  
The masers have a projected separation of 2 AU with the position angles of the tangential velocities 
ranging from 196\deg\ to 242\deg, 
implying an opening angle of the flow of $\sim$45\deg.  
The separating motion of the masers perpendicular to the outflow axis of about 3.5 km s$^{-1}$ can be most simply explained by the
opening angle of the wind.    
The average position angle of 220\deg\ is closely aligned with the position angle of the giant HH flow.
Their large tangential velocities relative to the radial velocities (Table 4) suggest the flow lies
primarily in the plane of the sky.  This geometry also explains the apparent 
presence of red- and blue-shifted masers in the same outflow lobe which has also been
observed in the NGC~1333-SVS~13 outflow (Wootten \etal 2002).
While uncertainties in the radial and tangential velocities could place all three maser spots on the
either the redshifted or blueshifted side, the low inclination angle and wide-angle opening angle
of the flow are likely to produce both blueshifted and redshifted 
gas in the same outflow lobe (e.g., Cabrit \& Bertout 1986, 1990; Lee et al. 2000). 
An outflow orientation in the plane of the sky is commonly observed in other
low mass YSOs with maser emission and is expected if the maser amplification
occurs in velocity coherent planar shocks (Claussen \etal 1996).
The relationship of the masers to known outflows in the region is discussed below.

\section{Discussion}

It is clear from our maser observations and those of the HH/radio continuum jet, that all 
of the reported outflow activity originates from the vicinity of Haro~6-10S.  
Apart from H$_2$ emission in the spectrum of Haro~6-10N,
there is not as yet evidence for an outflow directly related to this YSO.

It is also apparent that there are two distinct outflows originating from Haro~6-10S.
Both the giant HH flow (HH411/412/410, HH184~A-E) and the water masers align along a position angle of $\sim$225\deg\
and indicate an episodic outflow nearly in the plane of the sky.
Working from furthest out to closest in,  HH411/410 have dynamical ages
of 6000 years; HH184~E an age of 1000 years;  HH184~B-C an age of 750 years; and HH184~A an age of 
 250 years.\footnote{A velocity of 
$\sim$150 km s$^{-1}$ was assumed, e.g., Schwartz, Jones \& Sirk 1984.}
Finally, the water masers have a dynamical age estimated from the space velocities given in Table
4 and the distance from the 3.6 cm continuum radio emission of $\sim$1 year.  
The opening angle of 45\deg\ for the maser outflow is consistent with the collimation of the outflow 
increasing as the dynamical age decreases
(Arce \etal 2007).  
The episodic nature of HH flows has been
discussed by Reipurth (2000) in terms of interactions between components of binary
or triple systems and their associated disks.  Alternatively, magnetorotational instabilities
can lead to rapid accretion events without invoking a close companion (Zhu \etal 2009).
It is worth noting that the maser space velocities we measure are somewhat lower than those of
water masers near driving sources (20 - 40 km s$^{-1}$ e.g., Claussen et al.
1998; Moscadelli et al. 2006; Hirota et al. 2008; Marvel et al. 2008) of low-mass YSOs
and lower than observed from larger scale HH proper motions.  
It is not unexpected that space velocities of other manifestations of the outflow (e.g., HH objects) 
are faster than the water masers, since, for both wind-driven and jet bow-shock models, the velocities 
rise with increasing distance from the source (Arce et al. 2007 and references therein).

A second outflow is suggested by the highly blue-shifted (150 km s$^{-1}$) Br$\gamma$ emission 
observed from Haro~6-10S by Beck \etal  Such radial velocities are consistent an outflow highly inclined to
the plane of the sky and with a face-on disk geometry as proposed  
for Haro~6-10S by Roccatagliata et al. (2011).
We propose this second outflow is delineated by the radio continuum jet, the
compact Br$\gamma$ and forbidden line emission, and the HH objects HH184~F
and G.
The position angles measured for these observations all lie in the range of 
175\deg\ - 195\deg. 
It is important to note that the extended Br$\gamma$ emission
initially has a position angle of $\sim$180\deg\ before turning toward the southwest at a position
angle of $\sim$200\deg (Beck, Bary, \& MacGregor 2010). 

Having two outflows arising from very close to Haro~6-10S strongly suggests that the 
driving source is not a single star, but has at least one close-in companion.  We 
estimate that the companion must be at least within 150 mas (20 AU) of Haro~6-10S; 
otherwise, given the resolution of radio continuum observations (290 mas  x 260 mas, Reipurth et al. 2004),
a separate source might have been detected depending 
on its brightness. 
Indeed, the convergent points for masers A and C and masers B and C lie within 35 mas of the radio continuum position.
We speculate that 
the driving source for the giant HH/maser flow originates from the close-in companion which,
due to an edge-on disk, 
contributes little to the infrared flux from Haro~6-10S.   The visible T Tauri star with a face-on disk may be
responsible for the highly blue-shifted Br$\gamma$ emission. 
High sensitivity, higher angular resolution observations should be
attempted to resolve the putative two stars (this would likely require high frequency 
radio observations).  Our suggestion of a binary nature for Haro~6-10S is unrelated
to that suggested by Reipurth et al. (2004), since that binary companion is invoked 
to drive an apparent east-west radio jet.  It is also unlikely to be related to the companion
proposed by Doppmann et al. (2008), since the period and separation their proposed companion are much smaller
than we require.

\section{Summary}

We report an outburst of water maser activity from the young binary system Haro~6-10.  Using the VLBA over
5 epochs, we were able to track three masers and derive their absolute positions and
proper motions.
The masers represent one side of a bipolar outflow that lies nearly in the plane of the sky and 
are clearly associated with the southern component of the binary, the T Tauri star Haro~6-10S.
We describe the maser activity as the most recent manifestation of a time-varying bipolar outflow
that also includes the giant HH flow: HH~410, HH~411, HH~412, and HH~184A-E.  We attribute a jet seen
in Br$\gamma$, forbidden line, and radio continuum emission and HH~184F-G to a second outflow also clearly associated 
with the T Tauri star Haro~6-10S.  We propose that Haro~6-10S is itself a binary system with a projected separation of 
$<$150 mas and speculate that the maser activity originates from the component which is fainter in the infrared.
Despite the presence of H$_2$ emission in the spectrum of Haro~6-10N, we do not identify any of the
outflows/jets directly to this YSO.

\acknowledgments
We thank Kari Wojtkowski, Sean Brittain, and Matthew Troutman for assistance with the analysis of the
infrared echelle spectra of Haro~6-10.  We also thank Prof. Tetsuo Sasao for sharing his code
that estimates contributions to the proper motions from the solar motion and galactic rotation.
Bradley Gerling gratefully acknowledges a summer internship through the
NASA/Missouri Space Grant Consortium.

\pagebreak

\pagebreak

\begin{center}
{\bf Figure Captions}
\end{center}
\medskip
%
% Figure 1 -
%
\plotone{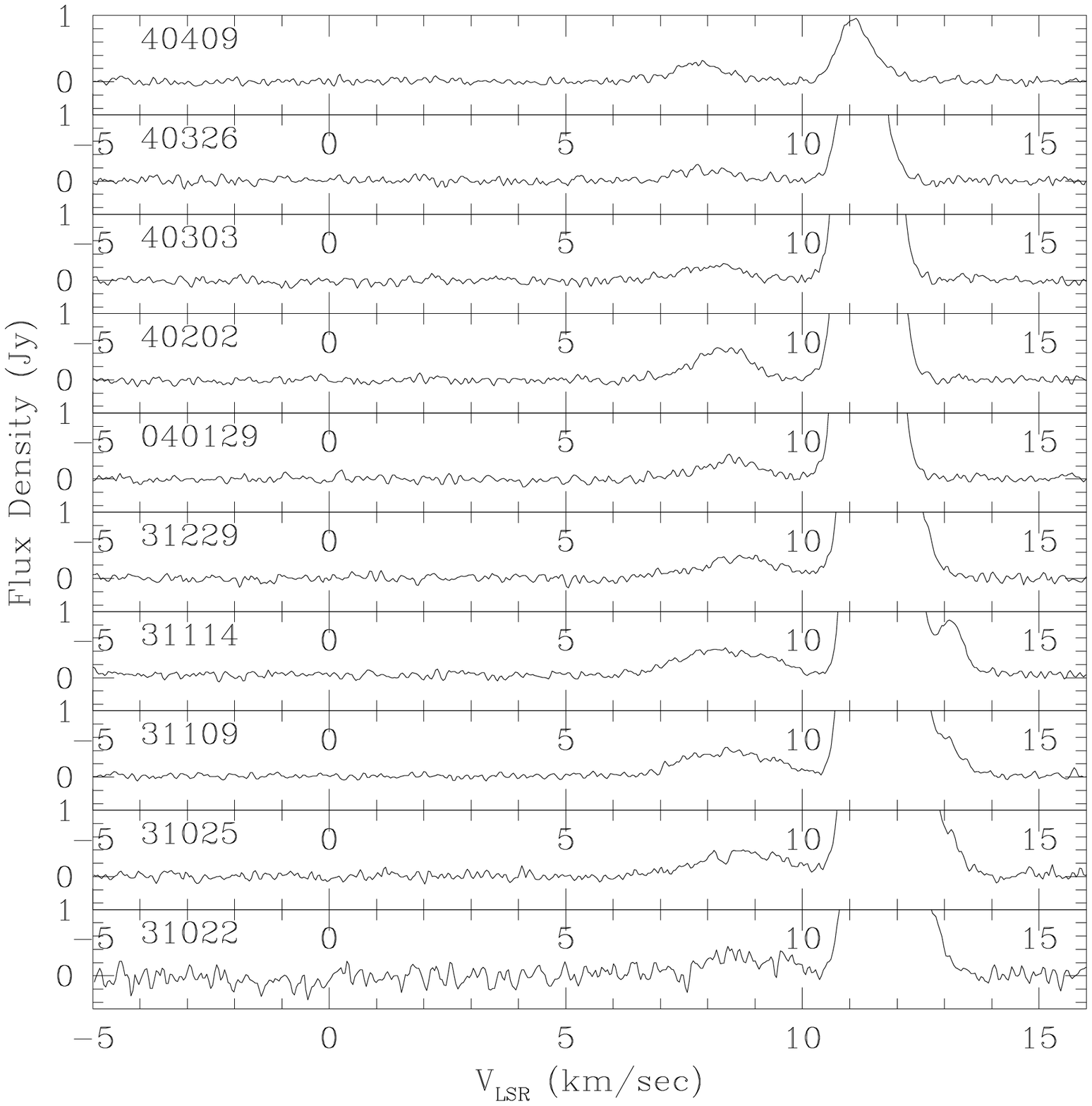}
\plotone{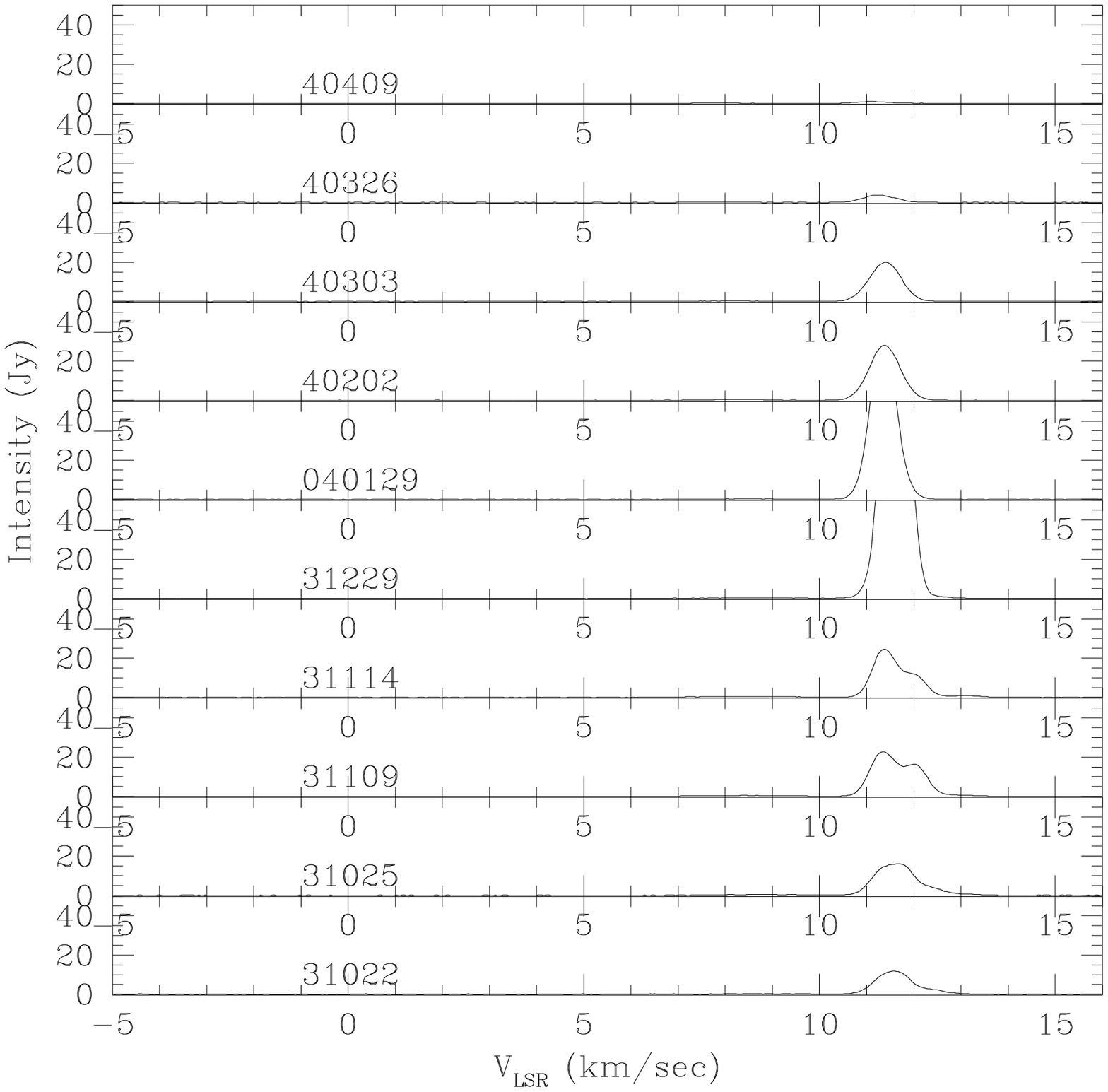}
\figcaption{%
GBT spectra of the Haro~6-10 masers.  Fig. 1a shows emission less than 1 Jy, focusing on the
weaker emission components.  Fig. 1b shows emission less than 50 Jy and  emphasizes the strongest
maser emission.
}
%
% Figure 2 -
%
\plotone{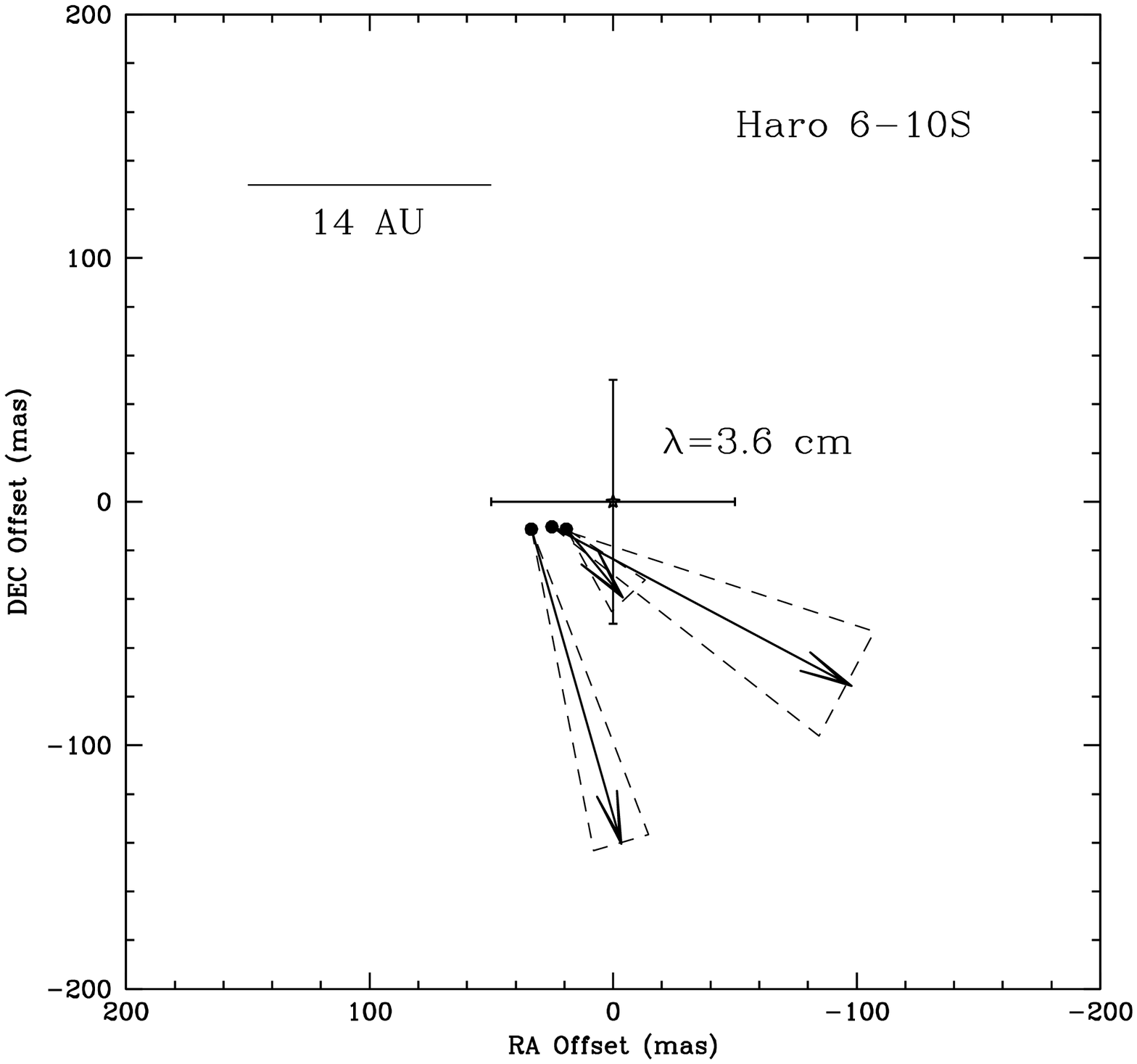}
\figcaption{%
Absolute motions for three maser components.  The (0,0) position is the radio continuum position of Haro~6-10S
(R.A.(2000) 4$^h$ 29$^m$ 23.733$^s$, decl.(2000.0) +24$^{\circ}$ 33' 00.12'', Reipurth \etal 2004). Solid arrows
mark the proper motion vectors, projected over a ten year period for display purposes.
Errors on the
proper motion magnitudes are given in Table 4.
The dashed lines delineate the uncertainties in the position angles.
}
%
%\documentclass[preprint]{aastex}
%\begin{document}
%\newcommand{\asec}{$^{\prime\prime }$}
%\newcommand{\amin}{$^\prime $}
%This is Table 2
\begin{deluxetable}{llcccccc}
\singlespace
%\rotate
\tablenum{2}
\tablewidth{0pt}
\tabletypesize{\scriptsize}
\tablecaption{Fitted Gaussian Components}
\tablehead{
\colhead{Epoch ID} & \colhead{Other} & \colhead{Julian Date} &  \colhead{X offset$\tablenotemark{a}$} 
& \colhead{Y offset} & \colhead{Radial Velocity} & \colhead{Peak Flux} & \colhead{FWHM}   \\
\colhead{} & &  & \colhead{(mas)} &  \colhead{(mas)} & \colhead{(km s$^{-1}$)} & \colhead{(Jy)} & \colhead{(km s$^{-1}$)} 
} 
\startdata
I-A   &       & 2453010.70 & 15.04 & -11.54 & 11.3  & 35   &  0.69  \\
I-B   &       &               & 14.16 & -12.01 & 11.2  & 6.2  &  0.60   \\
I-C   &       &               & 22.93 & -10.83 & 12.1  & 0.44 &  0.50  \\
I-D   &       &               & 13.36 & -11.46 & 11.3  & 0.69 &  0.65    \\
I-E   &       &               & 12.16 & -11.42 & 11.1  & 0.44 &  0.48   \\
      &       &               &       &        &   \\
II-A  & I-A   & 2453027.65 & 13.38 & -12.58 & 11.1  & 25   &  0.65   \\
II-B  &       &               & 19.47 & -10.97 & 11.0  & 0.52 &  0.49    \\
II-C  &       &               & 27.90 & -11.95 & 8.3  & 0.32 &  0.95  \\
II-D  &       &               & 13.76 & -12.32 & 11.2  & 8.5  &  0.86  \\
II-E  &       &               & 12.96 & -12.84 & 11.1  & 4.3  &  0.80   \\
      &       &               &       &        &   \\
III-A & II-A  & 2453043.61 & 12.44 & -13.28 & 11.1  & 35   & 0.62 \\
III-B & II-B  &               & 17.69 & -11.97 & 10.8  & 1.7  & 0.55 \\
III-C & II-C  &               & 27.16 & -13.08 & 8.5  & 0.10 & 0.85  \\
III-D &       &               & 11.89 & -12.61 & 11.2  & 2.6  & 0.60  \\
      &       &               &       &        &   \\
IV-A  &III-A  & 2453059.56 & 12.21 & -14.08 & 11.2  & 4.0  & 0.50  \\
IV-B  &III-B  &               & 17.27 & -12.85 & 11.0  & 0.81 & 0.40  \\
IV-C  &III-C  &               & 27.11 & -14.41 & 8.2  & 0.18 & 0.70 \\
IV-D  &       &               & 11.54 & -13.23 & 11.2  & 1.8  & 0.55 \\
IV-E  &       &               & 12.62 & -13.79 & 11.1  & 6.1  & 0.82 \\
IV-F  &       &               & 17.54 & -12.66 & 10.9  & 0.80 & 0.58 \\
      &       &               &       &        &   \\
V-A   &       & 2453073.50 & 12.53 & -14.46 & 11.5  & 5.1  & 0.59  \\
V-B   &       &               & 11.86 & -13.73 & 11.3  & 2.2  & 0.55  \\
V-C   &       &               & 11.49 & -13.48 & 11.1  & 2.0  & 0.70  \\
V-D   &       &               & 19.94 & -11.41 & 11.0  & 0.46  & 0.40  \\
V-E   &       &               & 18.34 & -13.09 & 10.9  & 0.25: & 0.55:  \\
V-F   &       &               & 17.71 & -13.12 & 10.9  & 0.66  & 0.69  \\
V-G   &       &               & 12.67 & -14.38 & 11.0  & 0.91  & 0.76  \\
\enddata
\tablenotetext{a}{Offsets are positive east and north from the $\lambda$ = 3.6 cm radio continuum source 
VLA~1 (RA(2000) = 4$^h$29$^m$23.733$^s$, DEC(2000) = +24$^{\circ}$33\amin 00.12\asec).  
Offsets in Epoch V are uncertain due to a possible 2-3 pixel shift in the reference maser position.}
\end{deluxetable}

%\end{document}

%\documentclass{aastex}
%\begin{document}
%This is new Table 3
\begin{deluxetable}{lccc}
\singlespace
%\rotate
\tablenum{3}
\tablewidth{0pt}
%\tabletypesize{\scriptsize}
\tablecaption{Corrections for Parallactic Modulation}
\tablehead{
\colhead{Epoch ID} & \colhead{Julian Date} &  \colhead{$\mu_{\alpha}$\tablenotemark{a}} 
& \colhead{$\mu_{\delta}$\tablenotemark{a}}    \\
\colhead{} &  & \colhead{(mas yr$^{-1}$)} &  \colhead{(mas yr$^{-1}$)}  
} 
\startdata
I    &     2453010.75 & -4.218  &  -0.379  \\
II   &     2453027.75 &  -5.711 &  -0.683  \\
III  &     2453043.75 &  -6.642 &  -0.913 \\
IV  &     2453059.75 &  -7.043 &  -1.069 \\
V   &     2453073.75 &  -6.940 &  -1.137\\
\enddata
\tablenotetext{a}{Corrections are to be subtracted from the values in Table 2.}
\end{deluxetable}

%\end{document}

%\documentclass{aastex}
%\begin{document}
\pagestyle{empty}
%\voffset=2.4truein
\begin{deluxetable}{lccc}
\singlespace
%\rotate
\tablenum{4}
\tablecolumns{4}
%\tabletypesize{\scriptsize}
\def\asec{$^{\prime\prime }$}
\def\deg{\hbox{$^\circ$}}
\def\amin{$^\prime $}
\tablecaption{Maser Proper Motions and Velocities}
\tablehead{
\colhead{Parameter} & \colhead{Maser I-A - IV-A} & \colhead{Maser II-B - IV-B} & \colhead{Maser II-C - IV-C}
}
\startdata
$\mu_{\alpha}$(mas yr$^{-1}$)											&   -0.10$\pm$1.17		&  -9.86$\pm$5.54		&  6.17$\pm$1.07			\\
$\mu_{\delta}$ (mas yr$^{-1}$)											&   -13.54$\pm$0.67		&  -17.10$\pm$0.35	&  -23.65$\pm$1.83		\\
$\mu_{\alpha,corr}$\tablenotemark{a}(mas yr$^{-1}$)		&    -2.62							&  -12.38						&  -3.65							\\
$\mu_{\delta,corr}$\tablenotemark{a}(mas yr$^{-1}$)		&    -2.83							&   -6.39						& -12.94							\\
$\mu_{total}$(mas yr$^{-1}$)												&     3.86							&   13.93						&  13.44							\\
V$_{tan}$(km s$^{-1}$)														&      2.6$\pm$0.6          & 9.3$\pm$3.3           & 9.0$\pm$1.2               \\
PA\tablenotemark{b}(degrees) 											&     223$\pm$14           & 242$\pm$10           & 196$\pm$5              \\
V$_{rad}$\tablenotemark{c}(km s$^{-1}$)							&     1.6                           &     1.4                      & -1.2                             \\
V$_{space}$(km s$^{-1}$)													&      3.0							&  9.4							& 9.1								\\
\enddata
\tablenotetext{a}{Proper motion correction for the solar motion.}
\tablenotetext{b}{Position angle in the plane of the sky measured east of north declination.}
\tablenotetext{c}{Radial velocity relative to that measured for Haro~6-10S in August 2003.}
\end{deluxetable}

%\end{document}

\end{document}